# Moderate-Density Parity-Check Codes

Samuel Ouzan and Yair Be'ery

Tel Aviv University, Ramat Aviv 69978, Israel

E-mail: {samouzan, ybeery}@eng.tau.ac.il

*Abstract*—**We propose a new type of short to moderate block-length, linear error-correcting codes, called moderate-density parity-check (MDPC) codes. The number of one's of the parity-check matrix of the codes presented is typically higher than the number of one's of the parity-check matrix of low-density parity-check (LDPC) codes. But, still lower than those of the parity-check matrix of classical block codes. The proposed MDPC codes are cyclic and are designed by constructing idempotents using cyclotomic cosets. The construction is simple and allows finding short block-length, high-rate codes with good minimum distance. Inspired by some recent iterative soft-input soft-output (SISO) decoders used in a context of classical block codes, we propose a low complexity, efficient, iterative decoder called Auto-Diversity (AD) decoder. AD decoder is based on belief propagation (BP) decoder and takes advantage of the fundamental property of automorphism group of the constructed cyclic code.**

## I. Introduction

Classical block codes [1] were the earliest type of codes to be discovered and to be used in practice. These codes are recently referred to as high-density parity-check (HDPC) codes [2]. In spite of the discovery of capacity-approaching codes [3] and the excellent tradeoff between performance and complexity that convolutional codes with Viterbi decoding achieve [4], algebraic block codes are still part of many industry standards error-correcting codes for high-rate, short-length applications, such as packet transmission, magnetic recording, thin information storage and compact discs. Indeed, they have a large minimum distance and can be decoded by several efficient soft decoding algorithms [5]-[7].

Meanwhile, because of the outstanding asymptotic performance of LDPC codes [8] under iterative BP decoding, some structured, moderate block-length, cyclic and quasi-cyclic LDPC codes [9]-[12] were designed. Moderate block-length structured LDPC codes are easy to implement and achieve good performance. However, for short block-length codes, it is difficult to construct a broad range of high-rate structured LDPC codes [29]. Furthermore, in that case, there is a tradeoff between designing a sparse parity-check matrix and keeping good minimum distance property [10].

By nature, algebraic block codes have a very dense parity-check matrix and the performance implied by straight-forward iterative BP decoding is poor. However, a number of authors considered various adaptations of BP decoding for classical block codes, which produce better performance, than standard sum-product (SP) decoding algorithm, [13]-[17]. The general tools considered are, cycles reduction algorithm, redundant Tanner graph, minimum weight, adaptive, and multiple parity-check matrices.

Adaptive belief propagation (ABP) algorithm proposed by Jiang and Narayanan [2], and ABP-OSD decoding [32], which combines ABP decoding [2] and ordered statistics decoding (OSD) [6], are kinds of algorithms that optimize the matrix representation during the decoding process and involve Gaussian elimination in every sum-product iteration. If high complexity is accepted, some relatively short algebraic codes decoded by these techniques can offer nearly optimum performance.

Random redundant iterative decoding [13], multiple-bases belief-propagation decoding [14], and modified random redundant decoding [15], represent another category, iterative, SISO decoding algorithms that do not require Gaussian elimination. For several well known, relatively small, algebraic codes, the decoding performance of those algorithms can be shown to closely follow that of maximum-likelihood (ML) decoders. However, for moderate block-length codes and at some relatively high coding rate, the gap to the maximum-likelihood decoder remains significant [25].

In our paper, we propose an alternative, for high-rate, moderate block-length, HDPC codes, and high-rate short structured LDPC codes. As we mentioned, the construction of small structured LDPC codes implies a constraint on the density of the parity-check matrix. We relax this constraint by designing MDPC codes with a parity-check matrix that contains a moderate number of one's. In a certain sense, and from a different perspective, we try to design codes that are more appropriate to low complexity, practical iterative decoders, used in the context of classical block codes [13]-[15].

We design ($n$, $k$, $d$) binary cyclic MDPC codes by constructing parity-check polynomial, obtained directly from idempotents via cyclotomic cosets mod $n$. In order to design a low complexity encoding/decoding scheme with possible practical utilizations, we propose a low complexity SISO diversity decoder called Auto-Diversity (AD) decoder. AD decoder belongs to the same family of low complexity iterative algorithms used for HDPC codes. Furthermore, AD decoder employs only a small number of redundant parity-checks and tries to minimize the number of operations not included in the regular BP algorithm.

We demonstrated that for memoryless additive white Gaussian noise (AWGN) channel, with low complexity iterative decoders used for HDPC codes and for a given, moderate block-length, high multiple coding rates, MDPC codes outperform equivalent Bose-Chaudhuri-Hocquenghem (BCH) codes [1], [19].

The remainder of this paper is organized as follows. In section II, we provide necessary definitions regarding cyclic codes, idempotents and automorphism group of cyclic codes. In section III, we introduce the construction method of MDPC codes. Illustration of the AD algorithm is proposed in section IV. Section V presents simulation result and complexity analysis. Finally, section VI concludes the paper.

## II. PRELIMINARIES

In this section we provide elementary definitions regarding idempotents of cyclic codes [1], [18]-[19]. This background is necessary in order to present the searching algorithm of MDPC codes. Furthermore, we give a brief description of automorphism group of cyclic codes; Essential property, which the decoder developed in this paper, is based on.

### A. Idempotents of Cyclic Codes

A binary cyclic code can also be represented as an ideal $I$, of the Ring

$$R_n = F_2[x]/(x^n + 1), \tag{1}$$

that consists of the residue classes of all polynomials with coefficients from the binary field $F_2$ mod $x^n + 1$. For every ideal $I$, there is a unique monic polynomial $g(x) \in I$ of minimum degree which divides $x^n + 1$, such that $g(x)$ generate the ideal. If $g(x)$ has degree $r$, the dimension of the code $k$ is $n-r$. Since $g(x)$ divides $x^n + 1$, there are $n-k$ zeros of $g(x)$ and all of them are $n$th root of unity. From the coefficients of $g(x) = g_0 + g_1 x + ... + g_r x^r$, the generator matrix can be expressed as

$$G = \begin{bmatrix} g_0 & g_1 & g_2 & \cdots & g_r & & 0 \\ & g_0 & g_1 & \cdots & g_{r-1} & g_r & \\ & & & \cdots & \cdots & & \\ 0 & & & g_0 & \cdots & \cdots & g_r \end{bmatrix}, \tag{2}$$

The parity-check polynomial $h(x)$ is determined by $h(x) = x^n + 1/g(x)$. From the coefficients of $h(x) = h_0 x + h_1 x + ... + h_k x^k$, the parity-check matrix can be expressed as

$$H = \begin{bmatrix} 0 & h_k & \cdots & h_2 & h_1 & h_0 \\ \cdots & h_k & h_{k-1} & \cdots & h_1 & h_0 \\ & & \cdots & \cdots & & \\ h_k & \cdots & h_1 & h_0 & & 0 \end{bmatrix}. \tag{3}$$

A polynomial $E(x)$ of $R_n$ is an idempotent if

$$E(x) = E(x)^2 = E(x^2). \tag{4}$$

A cyclic code $C$ of length $n$ has a unique idempotent $i(x) \in R_n$ called the generating idempotent such as

$$C = \langle i(x) \rangle, \tag{5}$$

where $\langle i(x) \rangle$ represents the principal ideal and consists of all multiples of a fixed polynomial $i(x)$ by elements of $R_n$.

Let $n$ be a positive odd integer and let $s$ be an integer in the interval $0 \leq s \leq n$. The 2-cyclotomic coset of $s \bmod n$ is the set

$$C_s = \{s2^j \bmod n \mid j = 0, 1, \ldots m_s - 1\}, \tag{6}$$

where $s$ is called the coset representative, which represents the smallest number in the coset, and $m_s$ is the smallest positive integer such that $s2^{m_s} \equiv s \bmod n$. From (6), two different cyclotomic cosets mod $n$, $C_s$ and $C_t$, are disjoint and the union of all the cyclotomic cosets mod $n$ represent the entire set, $\{0, 1, 2, \ldots, n-1\}$, since $n$ is odd and $C_s = C_t$ if and only if $t \in C_s$.

It is known that any idempotent $i(x)$ of $R_n$ generate some cyclic code of length $n$ and it can be expressed as

$$i(x) = \sum_{s \in S} \sum_{i \in C_s} x^i = i_0 + i_1 x + \ldots + i_{n-1} x^{n-1}, \tag{7}$$

where $S$ is a subset of the set $S^n$ which consists of all the representatives of the cyclotomic cosets mod $n$. When, the subset $S$ contains only one element, $i(x)$ is defined as a primitive idempotent. We can define the generating idempotent $i^{dual}(x)$ of the dual code $C^\perp = \{y \in (F_2)^n \mid xy^T = 0 \text{ for all } x \in C\}$ as

$$i^{dual}(x) = 1 + x^n i(x^{-1}). \tag{8}$$

Following the relation (2), the generator matrix $G$ of the cyclic code $C$ generating by $i(x)$ is the matrix with $k$ cyclic shifts of the coefficients of the polynomial $i(x)$. Hence, the parity-check matrix of the code $C$ generated by $i(x)$ is the matrix with $n$-$k$ cyclic shifts of the coefficients of the polynomial $i^{dual}(x)$.

## B. Automorphism group of cyclic codes

Let $C$ be a $(n, k, d)$ code. The Automorphism group of the code $C$, Aut($C$), is the set of permutations $S_n$ of coordinate places which send $C$ into itself, i.e. codewords go into possibly different codewords form the automorphism group of the code $C$.

$$\text{Aut}(C) := \{\pi \in S_n | \pi(C) = C\}. \tag{9}$$

If $C$ is a linear code and $C^\perp$ its dual, then Aut($C$)=Aut($C^\perp$). The automorphism group of a cyclic code contains all the cyclic permutations i.e., the cyclic permutation $(1, 2, ..., n-1)$ and all its powers [18]. Because $n$ is odd, the map $\sigma_2: x \to x^2$ is a permutation of $R_n$, therefore the automorphism group of a cyclic code is generated by the permutations

$$\pi^{(j)} = (1,\ 2^j + 1,\ 2 \cdot 2^j + 1, ...,\ (n-1) \cdot 2^j + 1),\ 0 \leq j \leq |C_1|, \tag{10}$$

where $|C_1|$ is the cardinality of the cyclotomic coset with coset representative one, defined in (6).

## III. Construction of the Moderate-Density Parity-Check Codes

In this section we provide a simple construction algorithm which allows finding MDPC codes. It consists of searching bounded weight idempotents, which will serve to construct moderate-density parity-check matrices. We implemented this algorithm using the software GAP [20] with the package GUAVA [21]. Algorithm 1 describes the proposed construction algorithm. Let $\Phi$ be the splitting field for $x^n + 1$ over $F_2$ and $\alpha \in \Phi$ be a primitive $n$th root of unity, the BCH bound of the code is determined by the number of consecutive powers of $\alpha$ mod $n$ which are roots of $g(x)$. The notations $|A|$ and $A[i]$ can either denote, if $A$ is a set, the number of elements and the $i$th element of $A$, whereas they represent, if $A$ is a superset, the number of sets and the $i$th set of $A$, respectively. Likewise, $A[i][j]$ represents the $j$th element of the $i$th set of the superset $A$. The notation $\text{GCD}(g(x), f(x))$ refers to the greatest common divisor of two polynomials $g(x)$ and $f(x)$. We denote $\theta$-$combination$ of a set $S$, an unordered subset of $\theta$ elements of $S$. We assume that all the operations are performed over the binary field $F_2$.

**Input :** $n \Leftarrow$ block-length of interest.
$d \Leftarrow$ minimal minimum distance of interset.
$N_{id} \Leftarrow$ number of primitives idempotents allowed.
$C_s \Leftarrow$ cyclotomic coset mod $n$ with coset representative $s$.
$S^n \Leftarrow$ set of all the representatives of the the cyclotomic cosets mod $n$.
$K \Leftarrow$ set of all $N_{id}$ - *combinations* from the set $S^n$-$\{0\}$, with cardinality $\dfrac{(|S^n|-1)!}{N_{id}!(|S^n|-N_{id}-1)!}$.

**Output :** $C \Leftarrow (n, k, d)$ linear cyclic code.

$i^{dual}(x) \Leftarrow 0;$
$R_{test} \Leftarrow 0;$
**for** $2 \leq j \leq |S^n|$
$\quad a_j(x) = 0;$
$\quad$ **for** $1 \leq k \leq |C_{S^n[j]}|$
$\quad\quad a_j(x) \Leftarrow a_j(x) + x^{n-C_{S^n[j]}[k]};$
$\quad$ **end for**
**end for**

**for** $1 \leq l \leq \dfrac{(|S^n|-1)!}{N_{id}!(|S^n|-N_{id}-1)!}.$
$\quad$ **for** $1 \leq m \leq N_{id}$
$\quad\quad i^{dual}(x) \Leftarrow i^{dual}(x) + a_{K[l][m]}(x);$
$\quad$ **end for**
$\quad i^{dual}(x) \Leftarrow i^{dual}(x) + 1;$
$\quad h(x) \Leftarrow \text{GCD}(i^{dual}(x), x^n + 1);$
$\quad g(x) \Leftarrow (x^n + 1) / h(x);$
$\quad r \Leftarrow$ degree of $h(x)$ ;
$\quad R \Leftarrow n / r;$
$\quad d_{bch} \Leftarrow$ bch bound of the code generated by $g(x);$
$\quad$ **if** $R \geq R_{test}$ AND $d_{bch} \geq d$ **then**
$\quad\quad C \Leftarrow$ replace by the cyclic code defined by $g(x);$
$\quad\quad i^{*dual}(x) \Leftarrow i^{dual}(x);$
$\quad\quad R_{test} \Leftarrow R;$
$\quad$ **end if**
$\quad i^{dual}(x) \Leftarrow 0$
**end for**

**Algorithm 1 :** Construction of MDPC codes.

As shown in section II-A, the parity-check matrix of the cyclic code provided by the algorithm can be designed as $n-k$ cyclic shifts of the coefficients of the polynomial $i^{*dual}(x)$. This construction algorithm permits to design several codes of same length, with different rates, by allowing, for each search a different number of primitive idempotents. In other words, by judiciously increasing or decreasing the density of the parity-check matrix, we can get codes with higher or lower rate. While controlling the density, this proceeding allows finding high-rate cyclic codes with good minimum distance properties. Some of these codes will be presented in section V (see Table 1).

## IV. AUTO-DIVERSITY DECODER

The presented Auto-Diversity (AD) decoder was inspired by recent research on low complexity iterative decoders for classical block codes [13]-[15]. The algorithm presented is not more sophisticate than other iterative algorithms applied on dense graphs. On the contrary, AD decoder aims to simplify this kind of algorithms in order to propose practical solutions for block codes with high coding rate. AD decoder uses only the automorphism group property and a small amount of redundant parity-checks of the cyclic codes to create diversity during the decoding process.

Let $C$ be a linear $(n, k, d)$ block code of length $n$, dimension $k$ and minimum distance $d$. Let $\boldsymbol{u} = [u_1, u_2, ... u_k]$ denote the binary vector information bits and $\boldsymbol{c} = [c_1, c_2, ... c_n]$ be the binary codeword. We assume that bits are modulated using BPSK (with 0 mapped to +1 and 1 mapped to -1) and transmitted over an AWGN channel. The real value vector $\boldsymbol{y} = [y_1, y_2, ... y_n]$ is used to denote the noisy receive word. The decoder input is the log-likelihood ratio (LLR) vector $\boldsymbol{L} = (2/\sigma^2) \boldsymbol{y}$, where $\sigma$ is the channel noise standard deviation on an AWGN channel.

The decoding algorithm presented is based on sum-product algorithm [22]. The Tanner graph TG($H$), [23]-[24], is redundant and is represented by a $m \times n$ parity-check matrix with $m > n-k$. However we try to minimize this redundancy and we design codes with $m \leq n/2$.

Algorithm 2 describes the proposed algorithm. The decoding algorithm starts to decode the length $n$ soft input **SI** vector using regular sum-product algorithm with a $m \times n$ redundant parity-check matrix $H$, which consist, when the decoder operates on MDPC codes, of $m$ cyclic shifts of the coefficients of the polynomial $i^{*dual}(x)$. If a valid codeword is reached at this stage, with a maximum number of iterations $I$, the decoder stops. Otherwise, the decoder stores the decoded codeword, and applies the sum-product algorithm on the same **SI** with another parity-check matrix $\sigma H$ obtained by permuting the column of $H$ from random elements of Aut($C$)[1]. The decoder repeats this process until the algorithm converges to a valid codeword. If after a maximum number of diversity stages $N_{ds}$, the algorithm did not converge to a valid codeword, the decoder estimates the codeword with a *least metric selector* (LMS) [14], [15].

**Input : SI** ⇐ Length $n$ soft input vector.
$I$ ⇐ maximum number of sum-product iterations.
$N_{ds}$ ⇐ maximum number of diversity stages.
**Output :** estimated codeword $\hat{c}$;
**for** $1 \leq i \leq N_{ds}$
  $\hat{c}_i$ ⇐ BP(**SI**, $H$, $I$ itterations);
  **if** $\hat{c}_i \cdot H^T = 0$ **then**
    $\hat{c} \Leftarrow \hat{c}_i$;
    **return** $\hat{c}$ and **stop**
  **else**
    $\sigma$ ⇐ random element of Aut($C$);
    $H \Leftarrow \sigma H$;
  **end if**
**end for**
$\hat{c} \Leftarrow \mathbf{argmax}_{s \in \{1...N_{ds}\}} \sum_{v=1}^{n} |y_v - \hat{c}_{s,v}|$

**Algorithm 2 :** Auto-Diversity decoder.

V. PERFORMANCE AND COMPLEXITY OF MODERATE-DENSITY PARITY-CHECK CODES

As we mentioned in the introduction, MDPC codes propose an alternative, for high-rate, moderate block-length, HDPC codes, and high-rate short structured LDPC codes. Therefore, we propose to compare high-rate BCH(127,$k$) codes where $92 \leq k \leq 106$ ($0.7244 \leq R \leq 0.8346$), with equivalent MDPC codes as illustrated in Table 1, constructed using Algorithm 1. On one hand, there is no equivalent length and rate structured LDPC codes reported in the literature. On the other hand, HDPC decoding of BCH codes with the same dimensions do not achieve such excellent performance compared to smaller algebraic block codes [13-15].

---

[1] For practical implementation issue, instead of decoding the soft-input vector **SI** on Tanner graph TG($\sigma H$), $\sigma \in Aut(C)$, it is equivalent to decode with soft-input vector $\sigma^{-1}$**SI** on TG($H$).

TABLE I

PARAMETERS COMPARISON OF BCH (127, $\kappa$) CODES, $92 \leq k \leq 106$ WITH EQUIVALENT MDPC CODES

| BCH Code | $d_{\min}(C^{\perp})$ | Rate | MDPC Code | $d_{\min}(C^{\perp})$ | Rate | $i^{*dual}(x)$ |
|---|---|---|---|---|---|---|
| BCH(127,92,11) | 32 | 0.724 | MDPC(127,92,10) | 22 | 0.724 | $x^{126} + x^{125} + x^{123} + x^{122} + x^{119} + x^{117} + x^{112}$ $+ x^{111} + x^{107} + x^{97} + x^{95} + x^{94} + x^{87} + x^{67}$ $+ x^{63} + x^{61} + x^{56} + x^{47} + x^{28} + x^{14} + x^{7} + 1$ |
| BCH(127,99,9) | 44 | 0,779 | MDPC(129,100,8) | 29 | 0.775 | $x^{128} + x^{127} + x^{125} + x^{121} + x^{120} + x^{114} + x^{113}$ $+ x^{111} + x^{99} + x^{97} + x^{93} + x^{72} + x^{69} + x^{65}$ $+ x^{64} + x^{60} + x^{57} + x^{36} + x^{32} + x^{30} + x^{18}$ $+ x^{16} + x^{15} + x^{9} + x^{8} + x^{4} + x^{2} + x + 1$ |
| BCH(127,106,7) | 48 | 0.834 | MDPC(127,106,6) | 36 | 0.834 | $x^{126} + x^{125} + x^{123} + x^{120} + x^{119} + x^{113} + x^{111}$ $+ x^{106} + x^{104} + x^{99} + x^{93} + x^{95} + x^{90} + x^{86}$ $+ x^{85} + x^{81} + x^{71} + x^{70} + x^{64} + x^{63} + x^{60}$ $+ x^{53} + x^{52} + x^{45} + x^{43} + x^{35} + x^{32} + x^{30}$ $+ x^{26} + x^{16} + x^{13} + x^{8} + x^{4} + x^{2} + x + 1$ |

Table 1 indicates that the minimum distance (determined using MAGMA [26]) of the MDPC codes presented are still inferior, but very close to the minimum distance of equivalent high-rate BCH codes Figs. 1-3 present error performance on AWGN channel in terms of bit error rate (BER), and frame error rate (FER), respectively. We compare simulation results with the union bound (UB) given by

$$\text{BER} \leq \frac{1}{n}\sum_{\delta=d}^{n} A_{\delta} \cdot \delta \cdot Q(\sqrt{2\frac{k}{n}\delta\frac{E_b}{N_0}}),$$

$$\text{FER} \leq \sum_{\delta=d}^{n} A_{\delta} \cdot Q(\sqrt{2\frac{k}{n}\delta\frac{E_b}{N_0}}),$$

(10)

where $A_{\delta}$ (computed using the software GAP [20] with the package GUAVA [21]) denotes the number of codewords of weight $\delta$. This bound is known to be tight to the ML decoder for error rates of interest [4]. Table 2 presents the first fifteen values of $A_{\delta}$. Notice that MDPC codes presented have relatively few lowest codewords.

*A. Performance Analysis*

As MDPC codes, equivalent BCH codes are decoded with AD algorithm. In order to construct the redundant parity-check matrix of the BCH code *C* employed by the decoder, the minimum codeword of the dual code $C^\perp$ (determined using MAGMA [26]) is cyclically shifted $m$ times.

TABLE II
WEIGTH DISTRIBUTION

| | BCH Codes | | | MDPC Codes | | |
|---|---|---|---|---|---|---|
| $\delta$ | $A_\delta(127,92)$ | $A_\delta(127,99)$ | $A_\delta(127,106)$ | $A_\delta(127,92)$ | $A_\delta(129,100)$ | $A_\delta(127,106)$ |
| 6 | | | | | | 2667 |
| 7 | | | 48387 | | | 49149 |
| 8 | | | 725805 | | 12900 | 672973 |
| 9 | | 62230 | 8249920 | | | 8556625 |
| 10 | | 734314 | 97349056 | 17780 | 968919 | 99758246 |
| 11 | 112014 | 8454390 | 1065157128 | 96012 | | 1060095162 |
| 12 | 1082802 | 81725770 | 10296518904 | 856996 | 98943516 | 10250410030 |
| 13 | 4992624 | 706987918 | 90631060800 | 6353683 | | 90688861802 |
| 14 | 40654224 | 5756901618 | 737995780800 | 48487457 | 7281183894 | 738539743136 |
| 15 | 343960323 | 43470567491 | 5564376646815 | 354775643 | | 5563741707013 |

Fig. 1 presents BER and FER results of MDPC and BCH codes of length 127, dimension 92 and rate 0.724. The decoder utilizes only $m=60$ parity-checks, and $I=50$ iterations. We can observe that despite the fact that the minimum distance of the BCH code is slightly higher, the union bound of the two codes are almost equivalent. Indeed, this is indicating that MDPC code constructed presents overall good properties. Fig. 1a shows that MDPC(127,92) with AD decoding outperforms the equivalent BCH code at BER=$10^{-4}$ by 0.9 dB. Note that we obtain BER=$10^{-5}$ with MDPC(127,92) code, with a maximum number of diversity stages $N_{ds}=30$, at $E_b/N_0=4.55$, which is only 0.4 dB worse than ML performance (union bound). The gain produced by the MDPC(127,92) code is significant and it derives from the fact that we design a code with a relatively high minimum distance, good weight distribution and a Tanner graph which is more than 30 percent sparser than that of the equivalent BCH code.

Fig. 2 presents BER and FER results of a BCH code of length 127, dimension 99 and rate 0.779, compared with a MDPC code of length 129, dimension 100 and rate 0.775. We construct a code with slightly different parameters, since we did not find a good MDPC code with exactly the same length and dimension. However, both codes are comparable. The decoder utilizes only $m=55$ parity-checks, and $I$=50 iterations. The union bounds of the two codes are very close. Fig. 2b shows that MDPC(129,100) code outperforms the equivalent BCH(127,99) code, at FER=$10^{-3}$, with AD decoding, by 0.75 dB. We observe that the difference in terms of gain produce by the AD decoder is much important at the early diversity stages and it decreases exponentially with the augmentation of the number of diversity stages.

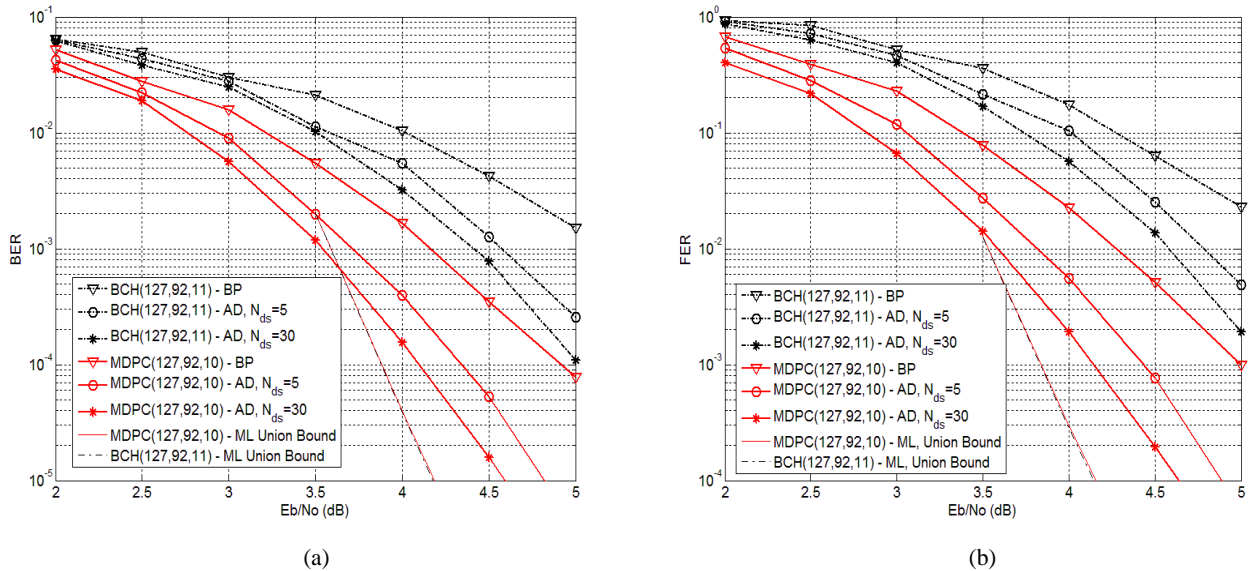

Fig. 1. BER (a) and FER (b) performance charts for MDPC(127, 92, 10) and BCH (127, 92, 11)

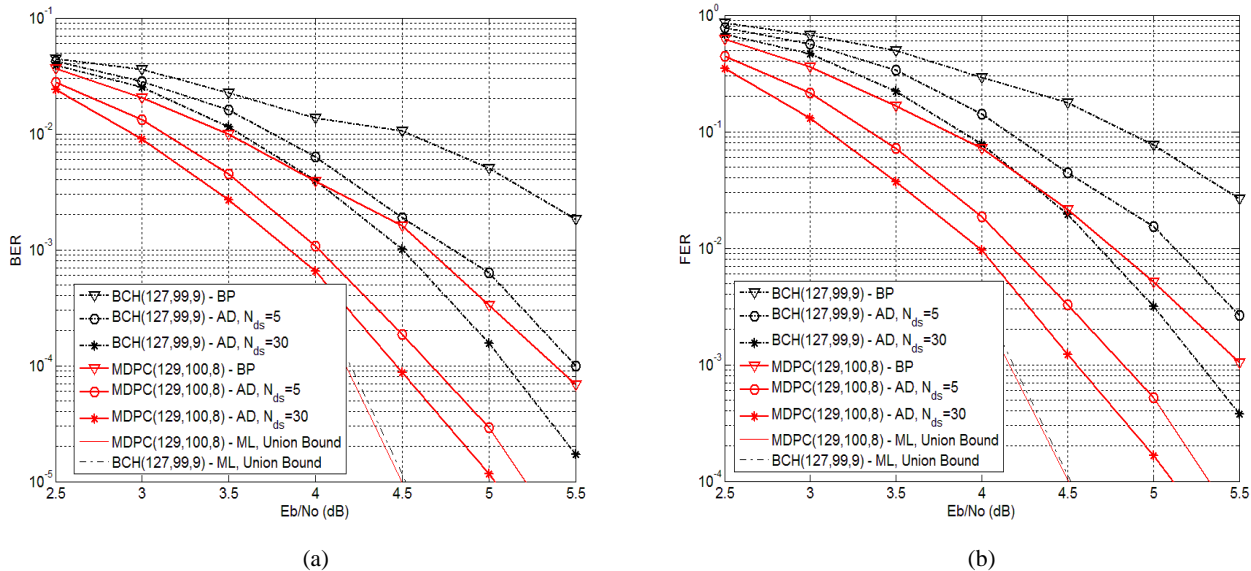

Fig. 2. BER (a) and FER (b) performance charts for MDPC(129, 100, 8) and BCH (127, 99, 9)

Finally, Fig. 3 presents BER and FER results of MDPC and BCH codes of length 127, dimension 106 and rate 0.834. The decoder utilizes only $m = 45$ parity-checks, and $I = 50$ iterations. As previous performance presented, MDPC code outperforms equivalent BCH code and performs close to the union bound. However, AD decoding algorithm for BCH(127,106) performs better than other BCH codes presented. It seems, from here and from previous work [14], that for high-rate codes, the behaviour of iterative decoding on dense graphs is better than for relatively lower rate codes. Note that each FER point was simulated until at least 100 frame errors were observed.

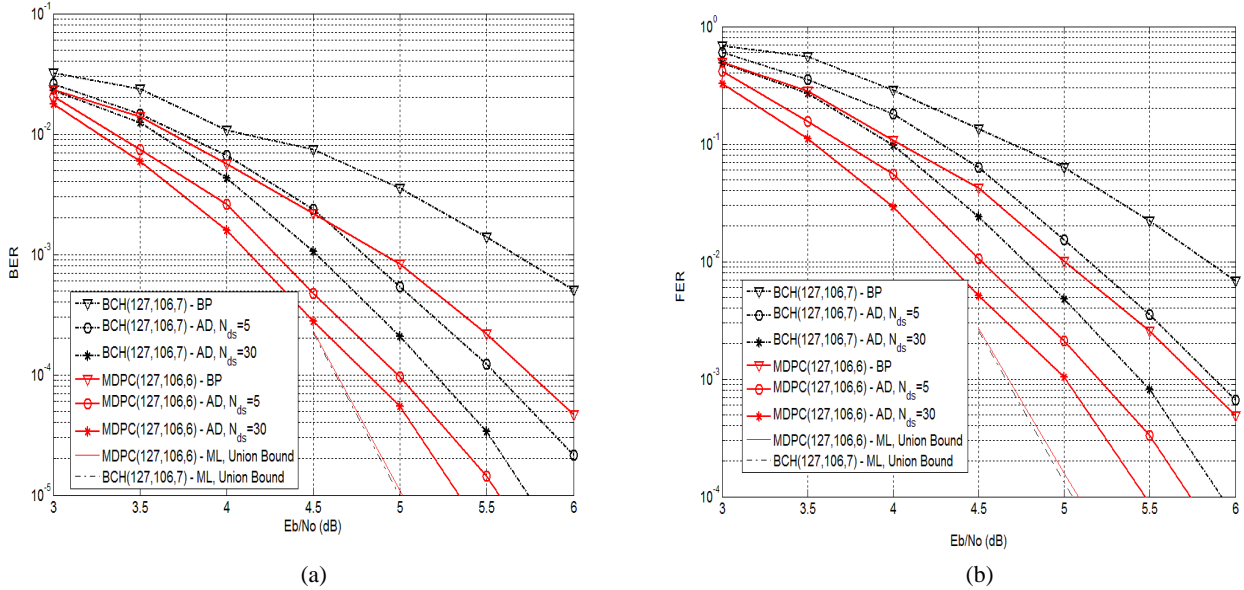

Fig. 3. BER (a) and FER (b) performance charts for MDPC(127, 106, 6) and BCH (127, 106, 7)

## B. Complexity Analysis

Figs. 4-6 illustrate the average number of iterations required to decode our proposed MDPC and equivalent BCH codes. For low probability of error, AD decoding provides significant coding gain over BP decoding, with a small increase of the average number of iterations. Furthermore, each iteration utilizes a relatively small number of redundant parity-checks. We can also observe that for every proposed MDPC codes, less than 3 iterations on the average are sufficient to reach BER=$10^{-5}$.

The relative complexity per iteration is given in Table 3. The number of edges, in MDPC and equivalent BCH Tanner graph is given by $d_{\min}(C^{\perp}) \times m$. Therefore, we set the relative complexity of MDPC codes to 1 and the relative complexity of BCH codes as the ratio $d_{\min}(C^{\perp}_{BCH}) / d_{\min}(C^{\perp}_{MDPC})$, where $C_{BCH}$ and $C_{MDPC}$ represents equivalent BCH and MDPC codes. We compare the overall complexity of $C_{BCH}$ and $C_{MDPC}$ by multiplying the average number of sum-product iterations with the relative complexity.

TABLE III

RELATIVE COMPLEXITY PER ITERATION

|  | MDPC(127,92) | MDPC(129,100) | MDPC(127,106) | BCH(127,92) | BCH(127,99) | BCH(127,106) |
|---|---|---|---|---|---|---|
| Relative Complexity | 1 | 1 | 1 | 1.454 | 1.517 | 1.333 |

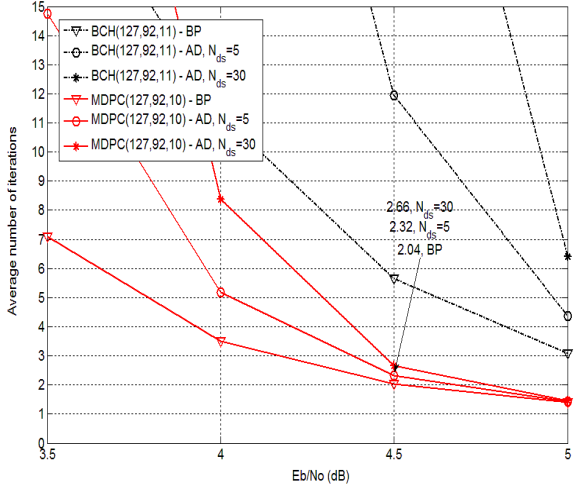

Fig. 4. Average number of iterations for MDPC(127,92) and BCH(127,92) code.

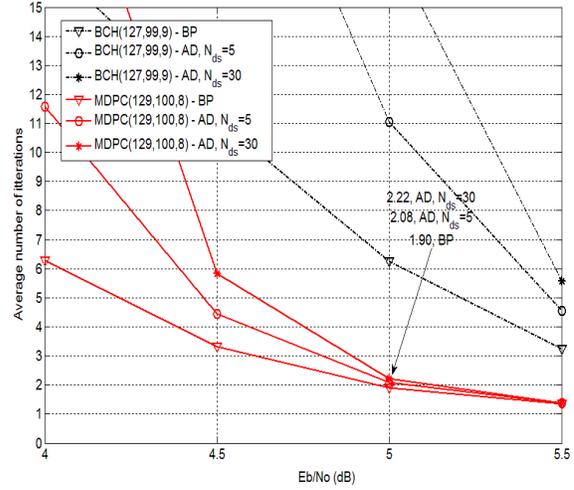

Fig. 5. Average number of iterations for MDPC(129,100) and BCH(127, 99) code.

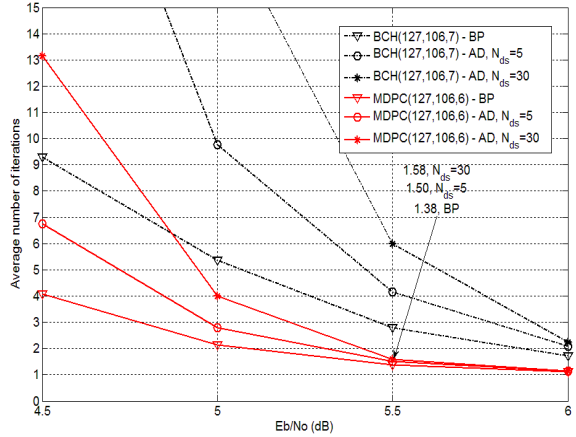

Fig. 6. Average number of iterations for MDPC(127,06) and BCH(127, 106) codes.

When $AD(m=5, N_{ds}=30)$ decoder is used, MDPC(127,106) and BCH(127,106) achieve FER=$10^{-3}$ at 5 dB and 5.45 dB respectively. One can observe from Fig. 3b, Fig. 6 and Table 3, that even though

BCH(127,106) requires an addition of 0.45 dB to achieve the same FER, the complexity of MDPC(127,106) is still less than half of the complexity of BCH(127,106). Keeping the same decoder parameters, similar comparison of complexity can be verified between other MDPC and BCH codes. For instance, from Fig. 2b, Fig. 5 and Table 3, MDPC(129,100) achieves FER=$10^{-3}$ with a reduction of complexity of about 60% and a coding gain of 0.75dB over BCH(127,99). As well as, form Fig 1b, Fig. 4 and Table 3, MDPC(127,92) achieves FER=$2\times 10^{-3}$ with a reduction of complexity of 10% but with a coding gain of 1 dB over BCH(127,92). Therefore, MDPC codes presented are more favourable than equivalent BCH codes with AD decoding in terms of performance and complexity. Note also that the proposed codes can serve as a good alternative to high-rate convolutional codes decoded by Viterbi algorithm [4].

## VI. Conclusion and Suggestions for Further Work

In this paper, a new kind of codes, called moderate-density parity-check codes has been presented. A simple construction algorithm based on the generation of idempotents using cyclotomic cosets mod *n*, has been demonstrated. Furthermore, we proposed a low complexity diversity decoder derived from iterative decoders applied for classical block codes. The codes presented are cyclic and can be encoded via a simple shift register [28]. We demonstrated that the proposed high-rates, MDPC codes significantly outperform equivalent high density BCH codes. Our main objective was to relax the tradeoff between designing high-rate and short block-length structured LDPC codes [29]. We achieved this task by, on one hand, increasing the density of the parity-check matrix and, on the other hand, applying iterative decoding technique suitable to decode codes induced by dense graphs [13]–[15]. The relaxation of density constraint also leads to design codes with larger minimum distance. MDPC codes could be suitable for many applications such as magnetic recording, optical communication and some delay sensitive services. High-rate is necessary to keep down the equalization loss and short length provides simpler system architecture [29], [30]. By relaxing somewhat the constraint on the low density, many families of iterative decodable codes such as LDPC codes based on finite geometries [31], or on balance incomplete block design (BIBD) [29] could be extend to MDPC codes.